\documentclass[CEJP,DVI]{cej} 

\usepackage{layout}
\usepackage{amsmath}
\usepackage{textcomp}

\usepackage{array}
\usepackage{rotating}
\usepackage{units}
\usepackage{amsthm}
\usepackage{amsmath}
\usepackage{graphicx}

\title{Numerical Analysis of Space Charge Effects in Electron Bunches at Laser-Driven Plasma Accelerators}

\articletype{Editorial}

\author{Anthony~Ashmore\inst{1},
        Riccardo~Bartolini\inst{1}\inst{2},
        Nicolas~Delerue\inst{1}\email{delerue@lal.in2p3.fr; Now at LAL, F-91898 Orsay, France}
}

\institute{
     \inst{1} John Adams Institute for Accelerator Science,\\
     University of Oxford, Keble Road, OX1 3RH Oxford, United Kingdom
     \inst{2} Diamond Light Source,\\
     Harwell, OX11 0DE Didcot, United Kingdom
          }

\abstract{
Laser-driven Plasma Accelerators (LPA) have successfully generated high energy, high charge electron bunches which can reach many kA peak 
current, over short distances. Space charge issues, even in transport lines as simple as a drift section, have to be carefully taken into account since 
they can degrade the beam quality, preventing any further application of such electron beams. We analyse the space charge 
effects within an electron bunch with numerical simulations in order to assess their effect on the beam. We use LPA 
beam parameters published in previous experimental studies.
These studies can give an indication of the working point where space charge can dominate the beam dynamics and has to be 
taken into account in the application of such beams.
}

\keywords{
Emittance \*\ Space-charge \*\ Laser-driven plasma acceleration.
}
\pacs{41.75.Jv, 41.75.Ht, 29.27.Bd, 29.27.Eg}

\begin{document}
\maketitle

\section{Introduction}

Laser-driven plasma wakefield accelerators have received much attention in the last years after the experimental 
demonstration of the possibility of plasma based acceleration~\cite{Dawson}, achieved at many different 
laboratories with the pioneering experiments in the early 2000s~\cite{nature}. Most notably, high-quality GeV 
electron beams were generated using cm-scale plasmas at Lawrence Berkeley National Laboratory~\cite{key-4}.
The ability to sustain extremely large acceleration gradients, enabling compact accelerating structures has opened 
the possibility of using such electron beams for a wide range of applications. Indeed, the low emittance 
and high peak current attainable has increased interest in laser-plasma acceleration as a driver for compact 
light sources~\cite{Jaro} ~\cite{Fuchs} and TeV-class linear colliders~\cite{Benedetti}. 
These applications, however, demand extremely good beam quality which has yet to be fully demonstrated 
experimentally, although rapid progress is expected from the new laser generation with multiterawatt pulses 
of a few femtoseconds length. 
While the low emittance and high peak current achievable with LPAs are necessary to drive Free Electron 
Lasers (FELs) they might also amplify space charge effects within the bunch, causing emittance growth and 
debunching~\cite{Gruner:2009zz} ~\cite{Geloni} limiting the possibility of using such beams.
Even the simple problem of transport of a high charge ultra short electron bunch requires careful 
analysis, since the bunch properties have to be maintained through the whole transfer line down to the final 
utilisation of the electron beam.
In view of the future application of LPA beams as a driver for FELs, this study aims to quantify to what extent the 
Coulomb forces within a bunch will affect the quality of the beam travelling in a drift section and suggest a 
boundary on beam parameters for a space charge dominated beam. The analysis presented here relies on an extensive 
campaign of numerical simulations performed with the code CSR-Track using the LPA beam parameters drawn from 
a survey of the main experimental data available to date. The analysis has also been extended to study the 
effect of space charge on LPA electron beams which constitute a modest improvement over the beam quality 
presently achieved by LPA.

\section{Numerical computation of space charge effects}

Laser-plasma acceleration is realized by using a short-pulse, high-intensity laser to ponderomotively drive
a large electron plasma wave (or wakefield) in an underdense plasma. The electron plasma wave has relativistic
phase velocity, approximately the group velocity of the laser, and can support large electric fields in the 
direction of propagation of the laser. As a result of this type of interaction, high peak current bunches can be 
generated with different beam parameters depending on the specific parameters of the laser and of the plasma used 
in the experiment. To investigate the effect of space charge in the propagation of these electron bunches we have
considered a simple drift space and followed the evolution of a macroparticle distribution along the drift as a 
function of the initial beam parameters such as charge, emittance and pulse length. The numerical analysis is based 
on the code CSR-track \cite{key-7} which is capable of describing the physics of beam transport in presence of 
space charge and coherent synchrotron radiation. The particle distributions used in these studies represent the 
beam qualities shown in Table 1. Where more than one set of parameters were reported for a given experiment, 
those which would produce the greatest space charge effects were used. 

\begin{figure}[htbp]
\begin{centering}
\includegraphics[width=100mm]{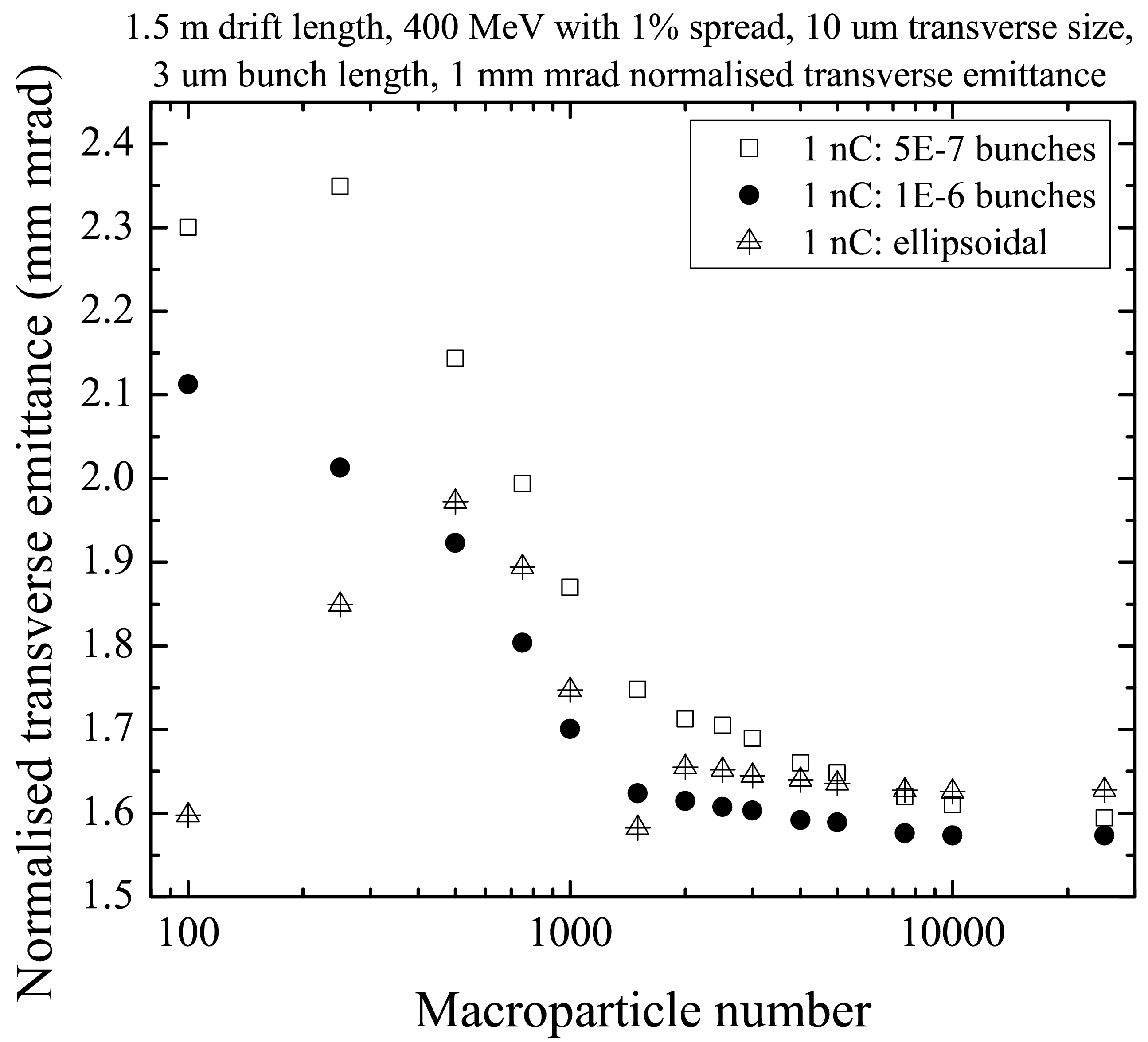}
\par\end{centering}

\caption{Calculated emittance increase as a function of the macro-particle number.}
\label{fig:macroparticle}
\end{figure}

CSR-track has several methods to compute the space charge effects. The method used in this report models the bunch 
as a collection of sub-bunches rather than point-like macroparticles and provides point to point calculations of 
Coulomb forces between sub-bunches. Shape and number of the sub-bunches were chosen as a compromise between speed 
and accuracy of the results. Clearly larger numbers of macro-particles will more accurately recreate the original 
bunch but lead to longer calculations: $3000$ macro-particles were used in this study, however, 
simulations ran with up to $25000$ macro-particles showed no significant change in results supporting this choice. 
In this study ellipsoidal self-scaling sub-bunches were used to cope with the significant expansion of the 
electron bunches under consideration. Fig. \ref{fig:macroparticle} shows the behaviour of emittance with varying 
macro-particle number for two fixed R.M.S. sub-bunch sizes and the self-scaling sub-bunches. A $\unit[1]{nC}$ 
bunch charge was used. For smaller charges these results are more stable at lower macro-particle numbers suggesting this is a worst 
case scenario which is not reached in the simulations carried out in this paper. The ellipsoidal self-scaling 
sub-bunches show the greatest stability against macro-particle number change above $ \sim 2000$ with a variation 
in emittance of $ \sim 1.5\%$ between $3000$ and $25000$. The bunches considered in this paper were drifted over 
2~m in vacuum unless otherwise specified.

\section{Results}

We report here the results of the numerical simulations perfomed with CSR-Track. We have used various 
initial beam distribution based on a number of examples of electron bunches produced in different experiments. 
In each case we have considered an electron bunch travelling in a drift section of $\unit[3]{m}$ and we have computed 
the evolution of emittance, energy spread and bunch length with the aim of understanding whether the beam properties are degraded as a result of the space charge self forces in the bunch. 
The analysis concentrates first on beam parameters already demonstrated in series of experiments carried out 
at different laboratories. Then we consider an example of an electron beam with properties which constitute 
a moderate improvement over the results presently achieved. The whole set of results are summarised in 
table~\ref{tab:summary}.

Kneip et al. \cite{key-2} reported the generation of a 200~MeV beam with 10~kA peak current and a normalised 
emittance estimated to 31.3~mm-mrad. The CSR-Track numerical simulations showed that the horizontal emittance of 
the bunch remained constant to less than $0.2\%$ suggesting the emittance growth due to Coulomb expansion
is negligible compared to the initial emittance. The transverse size of the bunch increased according to the 
initial divergence of the beam with no apparent space charge effects. The large divergence
caused the bunch to expand quickly in the transverse direction producing
a corresponding decrease in the strength of the Coulomb forces thus accounting
for the lack of significant space charge effects. Bunch lengthening in the longitudinal
direction was negligibly small due to the relatively large initial
bunch length which itself is Lorentz contracted. This also ensured
a stable peak current for the duration of the drift space.

Ibbotson et al. \cite{key-3} generated a 400~MeV beam with 0.875~kA peak current and a normalised emittance 
estimated to 34.4~mm-mrad. The CSR-Track numerical 
simulations showed that the horizontal emittance of the bunch remained constant to less than
$0.1\%$ which is within computational error. This suggests the bunch
is completely divergence-dominated with the Coulomb forces being negligible.
The transverse size increased exactly in line with that predicted
by a divergent, space charge free bunch model. This suggests that
space charge effects are negligible under these conditions. Bunch lengthening
was again negligible and the peak current constant.

Leemans et al. \cite{key-4} generated a 500~MeV beam with 12.5~kA peak current and a normalised emittance 
estimated to 24.5~mm-mrad. The CSR-Track numerical simulations showed that the horizontal 
emittance of the bunch remained constant to less than
$0.5\%$ which suggests the bunch is again divergence-dominated with
the Coulomb forces being negligible. The transverse size increased
in line with the divergence once again. The large initial emittance
and high bunch energy reduced the impact of any space charge effects. This result is expected when 
the beam parameters, including an increased beam energy, are compared with those of Kneip et al.

In the case of Rowlands-Rees et al. \cite{key-5} the parameters found in the paper would suggest a smaller 
emittance growth than those of Kneip et al. due to a smaller bunch charge. CSR-Track simulations with these 
parameters show no emittance growth with the transverse size being completely divergence dependent.

The beam paraemters considered so far, although achieved experimentally, do not yet provide electron 
bunches with the quality necessary to drive compact light sources based on FELs. We considered therefore the beam characteristics required for the Oxford Plasma Accelerator Light Source (OPALS) \cite{key-6} 
which is aimed at the production of FEL radiation in the VUV and Soft X-ray region. The predicted achievable 
emittance by the Laser Plasma Accelerator driving the OPALS FEL is much smaller than that examined in the other 
papers: the normalised emittance is 1~mm-mrad and the peak current 25~kA. The CSR-Track simulations show that 
for such small initial emittance the divergence driven expansion of the bunch is decreased. Due to this and the 
relatively large bunch charge we would expect significant space charge effects. Results for bunches of various 
energies with 25~kA peak current which evolve in a drfit over $\unit[1.5]{m}$ are found in table~\ref{tab:opals}.

\begin{table}
\begin{centering}
\begin{tabular}{cccccc}
 & \begin{sideways}
Kneip et al. \cite{key-2}%
\end{sideways} & \begin{sideways}
Ibbotson et al. \cite{key-3}%
\end{sideways} & \begin{sideways}
Leemans et al. \cite{key-4}%
\end{sideways} & \begin{sideways}
Rowlands-Rees et al. \cite{key-5}%
\end{sideways} & \begin{sideways}
Bajlekov et al. \cite{key-6}%
\end{sideways}\tabularnewline
\hline
\hline 
\noalign{\vskip\doublerulesep}
$\sigma_{z}$ ($\unit{fs}$) & $55$ & $80$ & $40$ & $45$ & $10$\tabularnewline[\doublerulesep]
\noalign{\vskip\doublerulesep}
$\sigma_{x}$ ($\unit{\mu m}$) & $22$ & $20$ & $25$ & $\sim 20$ & $10$\tabularnewline[\doublerulesep]
\noalign{\vskip\doublerulesep}
$\sigma_{x'}$ ($\unit{mrad}$) & $4$ & $2$ & $1$ & $ \sim 2$ & $\sim 0.13$\tabularnewline[\doublerulesep]
\noalign{\vskip\doublerulesep}
$Q$ ($\unit{pC}$) & $550$ & $70$ & $500$ & $100$ & $250$\tabularnewline[\doublerulesep]
\noalign{\vskip\doublerulesep}
$E$ ($\unit{MeV}$) & $200$ & $400$ & $500$ & $200$ & $400$\tabularnewline[\doublerulesep]
\noalign{\vskip\doublerulesep}
$E$ spread & $5\%$ & $5\%$ & $5\%$ & $\sim 5\%$ & $1\%$\tabularnewline[\doublerulesep]
\noalign{\vskip\doublerulesep}
$\epsilon_{n,rms}$ ($\unit{mm\: mrad}$) & $34.4$ & $31.3$ & $24.5$ & $\sim 15.7$ & $1$\tabularnewline[\doublerulesep]
\hline
\noalign{\vskip\doublerulesep}
$\sigma_{z}$ ($\unit{fs}$) & $120$ & $85$ & $41$ & $53$ & $10$\tabularnewline[\doublerulesep]
\noalign{\vskip\doublerulesep}
$\sigma_{x}$ ($\unit{\mu m}$) & $8000$ & $4000$ & $2000$ & $4000$ & $284$\tabularnewline[\doublerulesep]
\noalign{\vskip\doublerulesep}
$\epsilon_{n,rms}$ ($\unit{mm\: mrad}$) & $34.4$ & $31.3$ & $24.5$ & $15.7$ & $1.17$\tabularnewline[\doublerulesep]
\hline
\end{tabular}
\par\end{centering}

\caption{Simulation Parameters and their values after a $\unit[2]{m}$ drift}
\label{tab:summary}

\end{table}

\begin{table}
\begin{centering}
\begin{tabular}{ccccc}
 & \begin{sideways}
Initial value%
\end{sideways} & \begin{sideways}
$\unit[200]{MeV}$%
\end{sideways} & \begin{sideways}
$\unit[400]{MeV}$%
\end{sideways} & \begin{sideways}
$\unit[1]{GeV}$%
\end{sideways}\tabularnewline
\hline
\hline 
\noalign{\vskip\doublerulesep}
$\sigma_{z}$ ($\unit{fs}$) & $10$ & $10$ & $10$ & $10$\tabularnewline[\doublerulesep]
\noalign{\vskip\doublerulesep}
$\sigma_{x}$ ($\unit{\mu m}$) & $10$ & $382$ & $192$ & $77$\tabularnewline[\doublerulesep]
\noalign{\vskip\doublerulesep}
$E$ spread & $1\%$ & $1.38\%$ & $1.02\%$ & $1.00\%$\tabularnewline[\doublerulesep]
\noalign{\vskip\doublerulesep}
$\epsilon_{n,rms}$ ($\unit{mm\: mrad}$) & $1$ & $1.75$ & $1.10$ & $1.00$\tabularnewline[\doublerulesep]
\hline
\end{tabular}
\par\end{centering}

\caption{Change of beam parameters after $\unit[1.5]{m}$ drift space}
\label{tab:opals}

\end{table}

\begin{figure}[htbp]
\begin{centering}
\includegraphics[width=100mm]{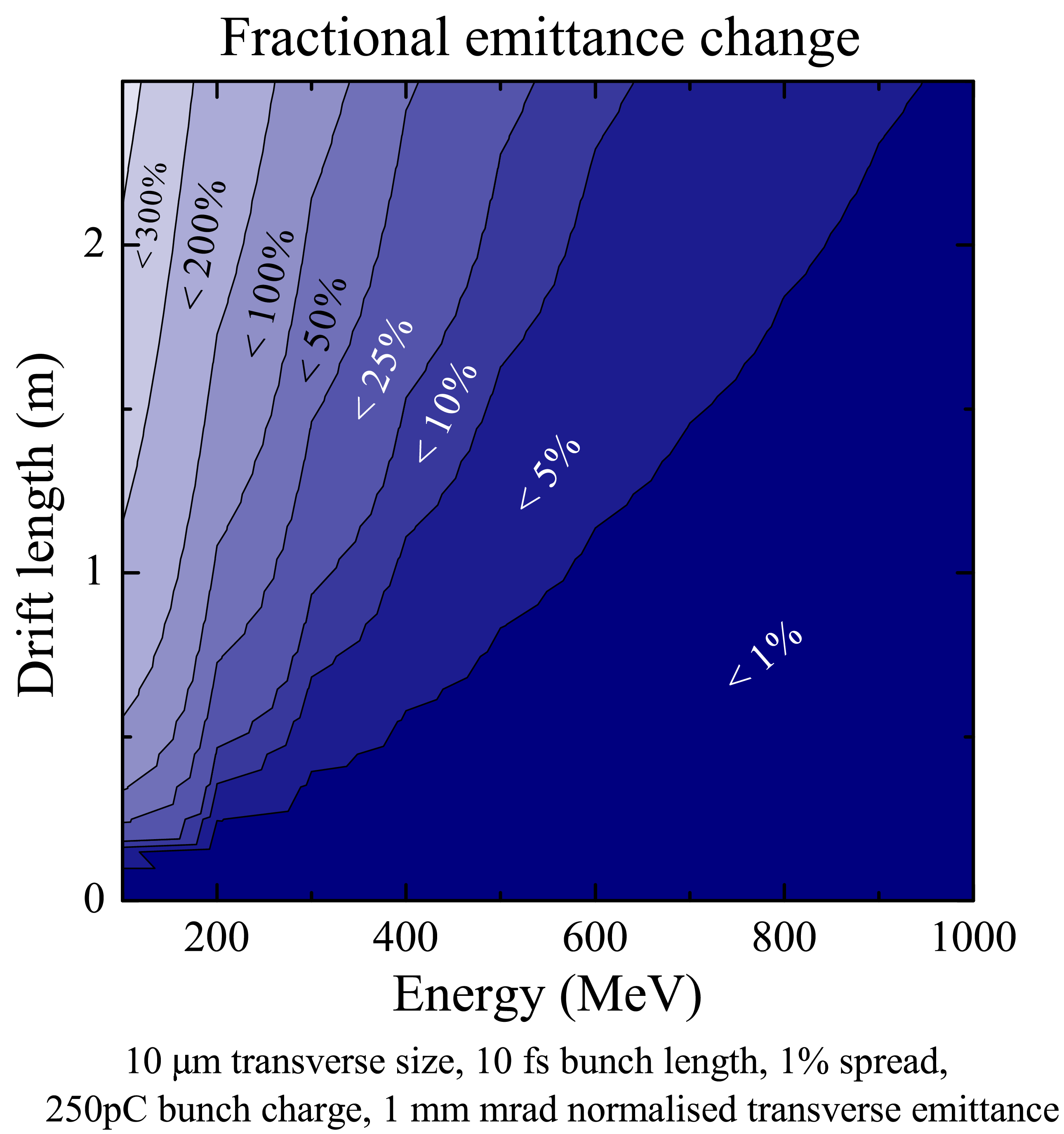}
\par\end{centering}

\caption{Contours of fractional emittance change as a function of beam energy and drift length.}
\label{Flo:contour_drift_energy}
\end{figure}

As the parameters of operation for OPALS are not yet fixed, we have performed an extensive parametric study 
of the initial beam characteristics on the final one after a drift section of 3~m. Contours plots of parameter 
space with varying bunch charge, energy and drift length were produced to investigate when space charge effects 
become important. The results can be seen in Fig. \ref{Flo:contour_drift_energy}.

At $\unit[400]{MeV}$ a $10\%$ increase in horizontal emittance is
reached after a $\unit[1.5]{m}$ drift length. The same space charge
effects have a much larger relative effect on the emittance compared
with the other papers examined due to both a smaller initial emittance
and the smaller divergence allowing the bunch to stay compact for
longer. At energies lower than $\unit[200]{MeV}$ it is apparent that
the emittance growth is very large even when drifted over relatively
small distances. To remain within a $1\%$ growth at $\unit[400]{MeV}$
the bunch can be drifted over approximately $\unit[0.6]{m}$. It is
also proposed to run OPALS in a more aggressive $\unit[1]{GeV}$ configuration;
this would allow much more freedom in how far the electron bunch can
be drifted without space charge adversely affecting its emittance.

\begin{figure}[htbp]
\begin{centering}
\includegraphics[width=100mm]{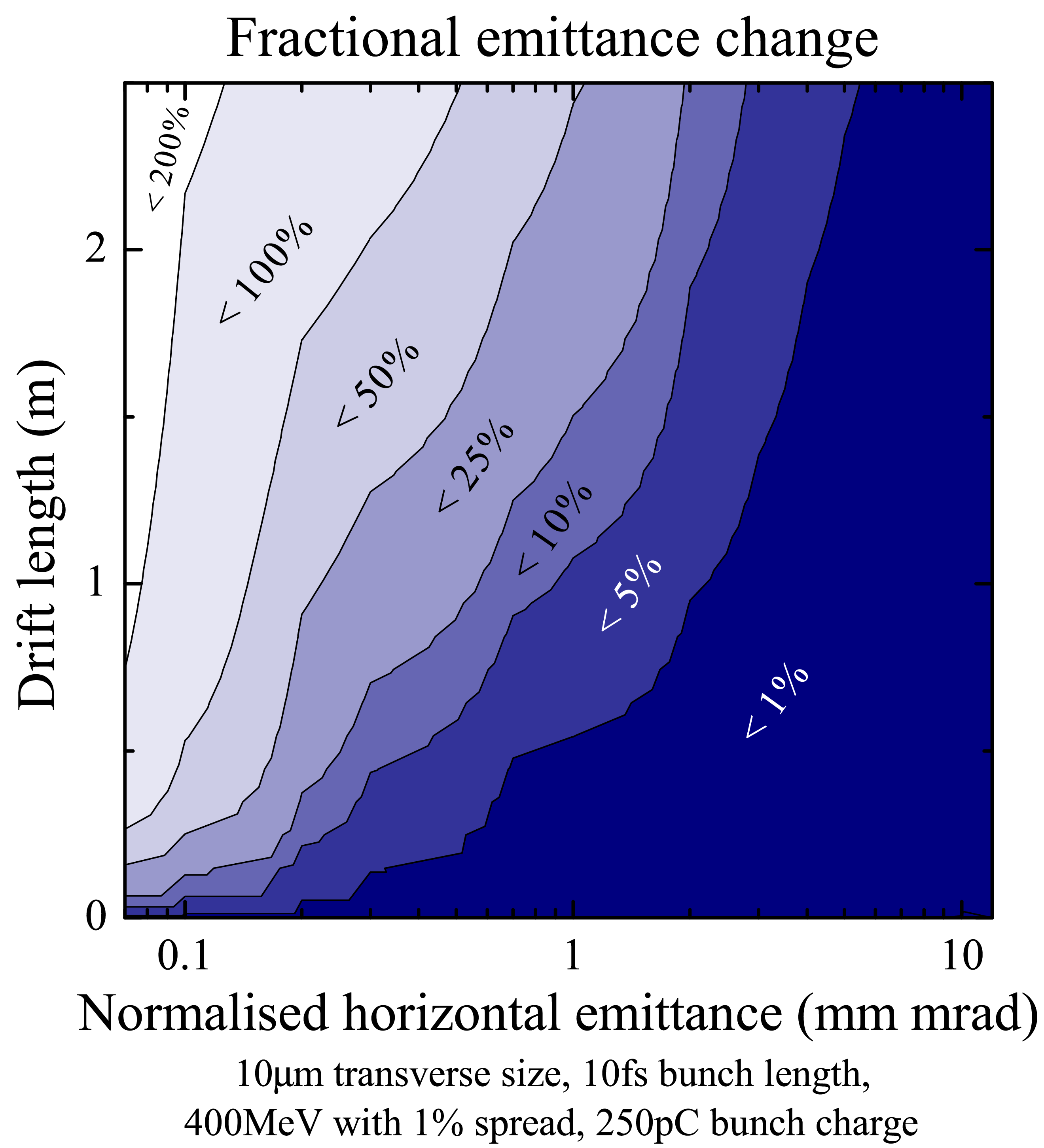}
\par\end{centering}

\caption{Contours of fractional emittance change as a function of initial emittance and drift length.}
\label{Flo:contour_drift_emit}
\end{figure}

To observe how the initial emittance of the electron bunch affected
the relative emittance growth, bunches with various emittances were
tracked over a $\unit[4]{m}$ drift length and the fractional change
calculated. As the bunch had a fixed initial size, varying the emittance
is equivalent to changing the initial divergence of the beam. The
results are plotted in Fig. \ref{Flo:contour_drift_emit}.

The experimental studies examined all have emittances above $\unit[10]{mm\: mrad}$
and exhibit the same small emittance growth as plotted. OPALS, with
its predicted $\unit[1]{mm\: mrad}$, can be expected to have an emittance
growth of more than $50\%$ if drifted for $\unit[4]{m}$ at $\unit[400]{MeV}$.
Looking to the future, with emittances $\sim \unit[0.1]{mm\: mrad}$
we can expect a doubling over $\unit[1.5]{m}$ unless the bunch energy
is increased. 

The variation of space charge forces with the bunch charge was also
examined and the effect on the emittance coupled with varying energy plotted
for an electron bunch drifted over $\unit[1.5]{m}$.

\begin{figure}[htbp]
\begin{centering}
\includegraphics[width=100mm]{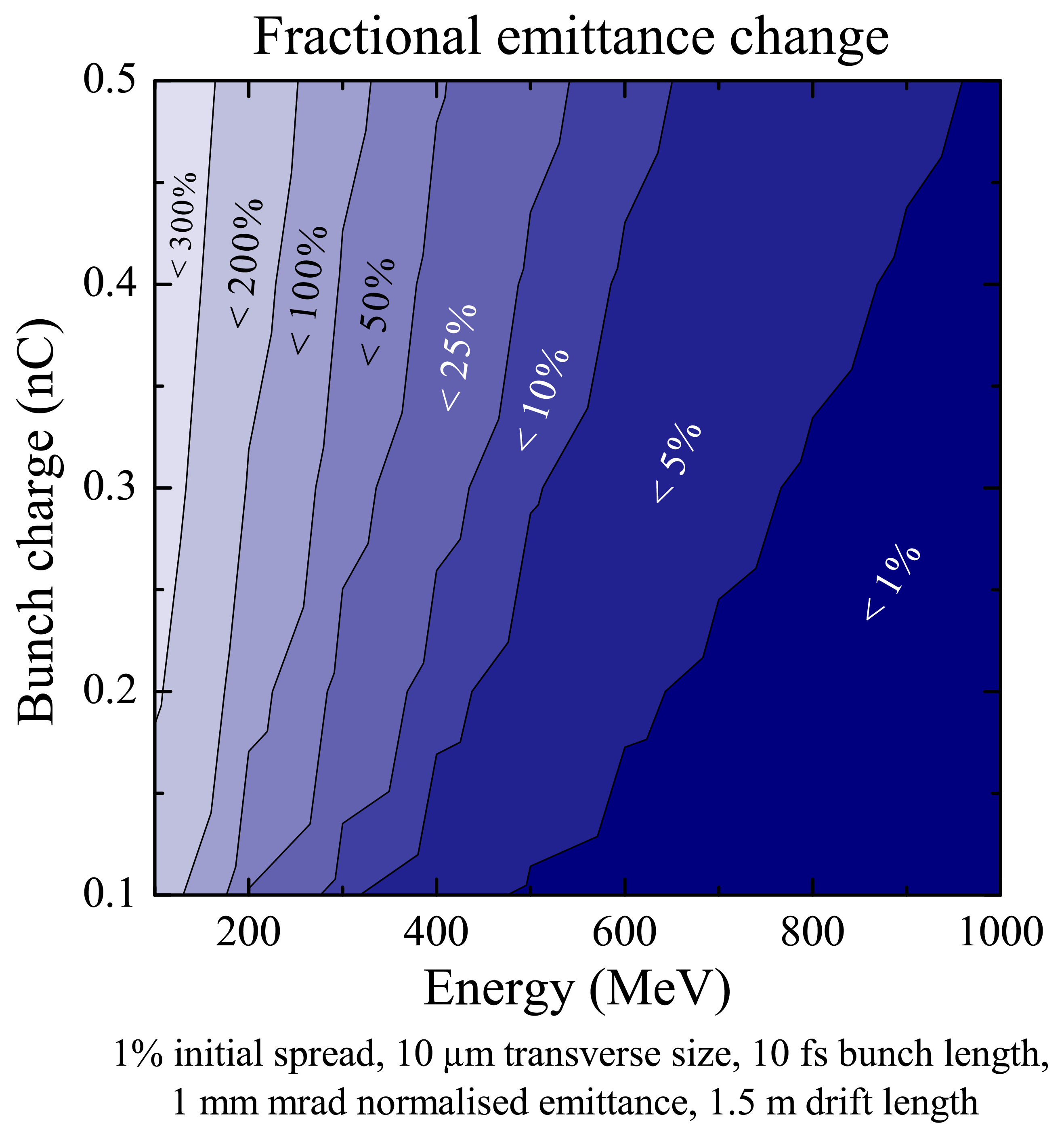}
\par\end{centering}

\caption{Contours of fractional emittance change as a function of beam energy and bunch charge.}
\label{Flo:contour_charge_energy}
\end{figure}

Fig. \ref{Flo:contour_charge_energy} shows the steep emittance growth
with increasing charge for low energy bunches confirming the unsuitable
nature of low energy, high charge bunches for applications requiring
a small emittance. The effect of the space charge forces drops off
rapidly with energy due to their $\gamma^{-2}$ dependence which originates
from the Lorentz contraction of the bunch observed in the lab frame.
At energies of $\unit[1]{GeV}$, as proposed in OPALS, bunch charges
of up to $\unit[0.5]{nC}$ could be drifted over $\unit[1.5]{m}$
with only a $1\%$ increase in transverse emittance. 

\begin{figure}[htbp]
\begin{centering}
\includegraphics[width=100mm]{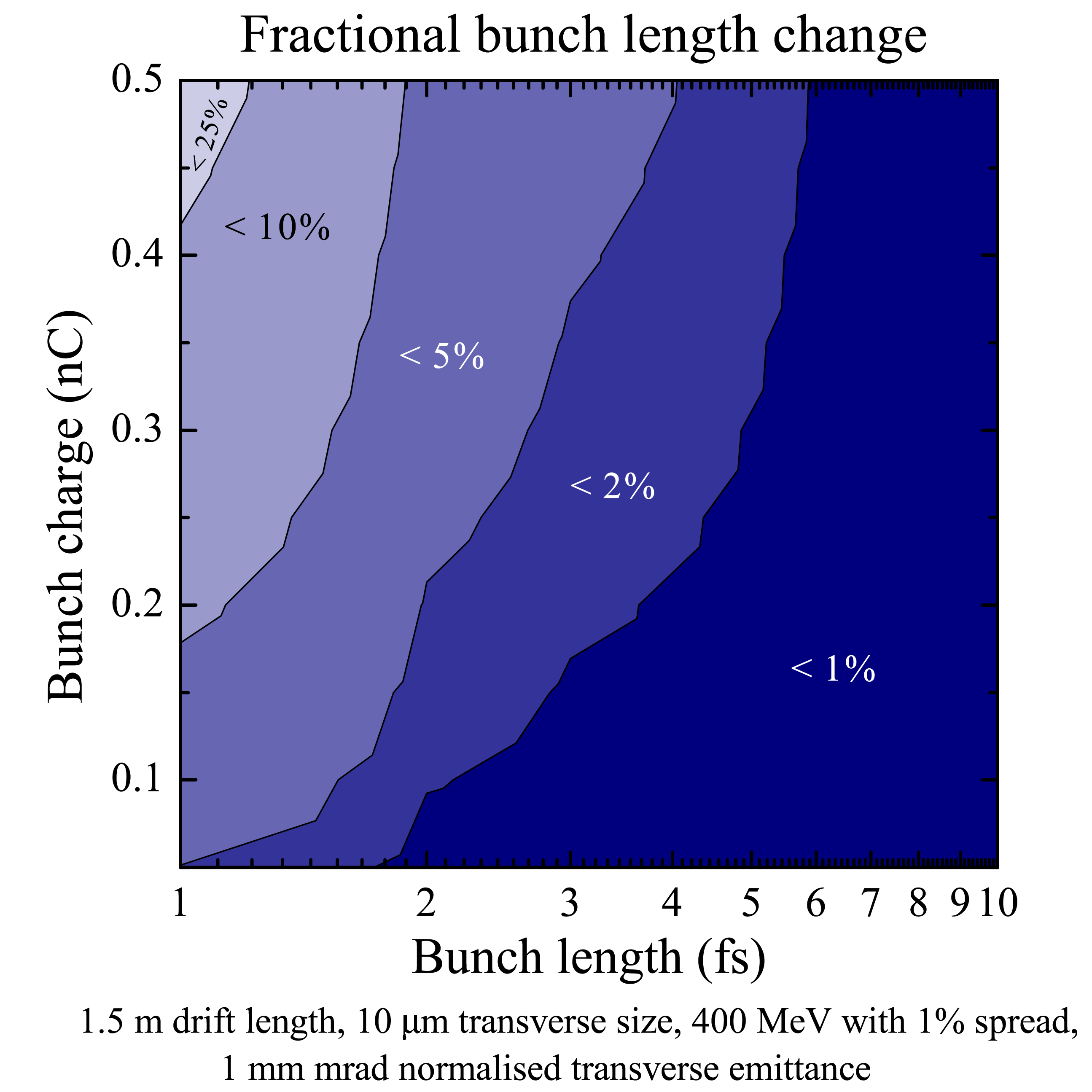}
\par\end{centering}

\caption{Contours of fractional bunch length increase as a function of initial bunch length and bunch charge.}
\label{Flo:contour_debunching}
\end{figure}

\begin{figure}[htbp]
\begin{centering}
\includegraphics[width=100mm]{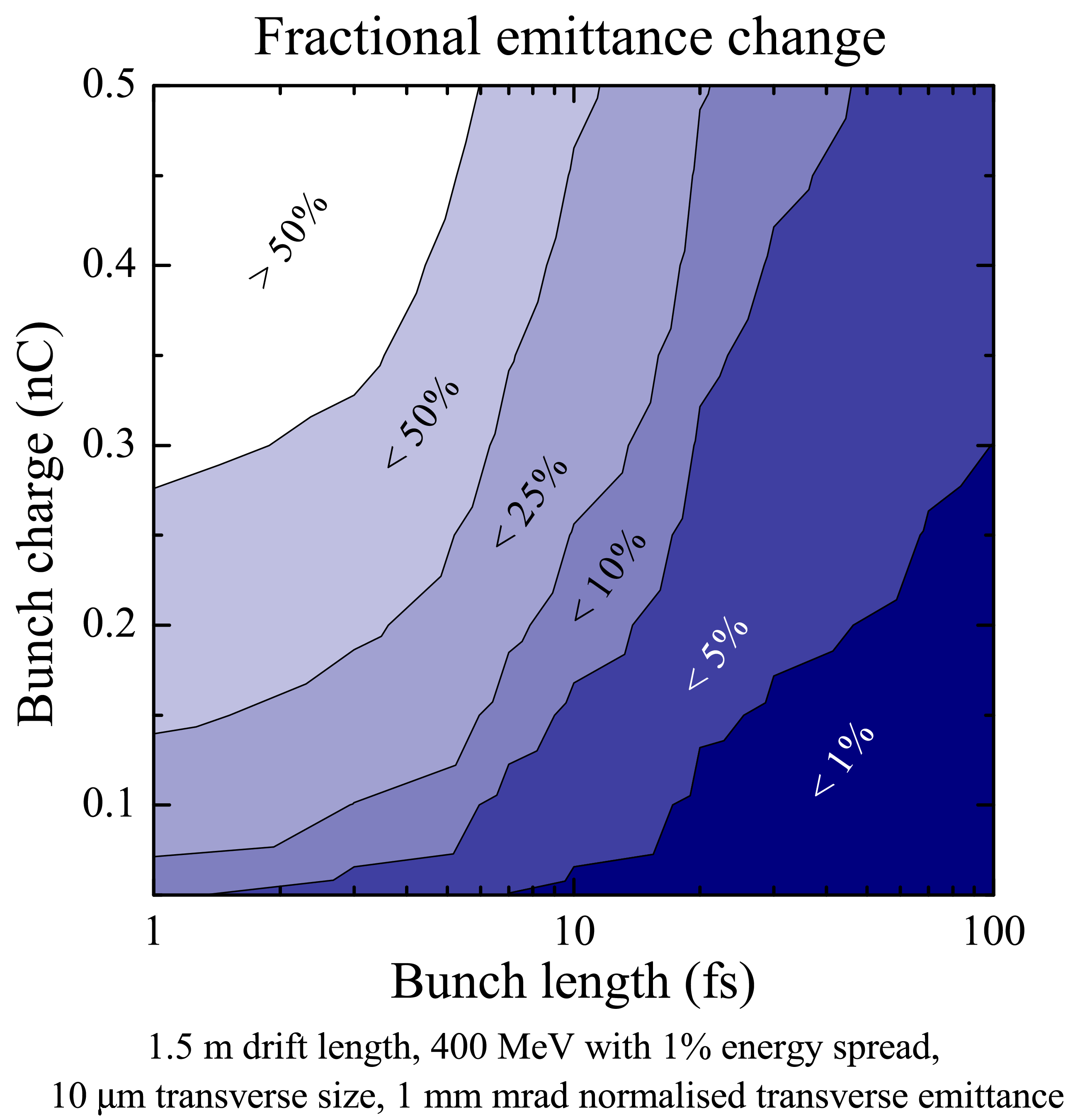}
\par\end{centering}

\caption{Contours of fractional emittance change as a function of bunch length and bunch charge.}
\label{Flo:contour_length_charge}
\end{figure}

A shorter bunch length will cause increased space charge forces. To
understand the magnitude of this effect, simulations were ran for
various bunch charges and lengths and the bunch length and emittance
increase calculated. The results are shown in Fig.'s \ref{Flo:contour_debunching}
and \ref{Flo:contour_length_charge}. At $\unit[10]{fs}$ and $\unit[250]{pC}$
the simulations showed less than a $1\%$ increase in the bunch length
suggesting debunching due to space charge is not of concern at OPALS.
At smaller bunch lengths and higher bunch charges more significant
effects are seen which may be of concern for future accelerators with
larger peak currents. 

Decreased bunch length at constant charge gives a larger peak current
leading to more Coulomb repulsion. For long bunches ($\sim \unit[100]{fs}$)
less than a $1\%$ increase in emittance would be seen at $\unit[250]{pC}$;
by $\unit[10]{fs}$ this is $10\%$ and by $\unit[1]{fs}$ is $45\%$.
The simulations outlined indicated how emittance growth will become a significant
problem if bunches become shorter without a corresponding increase
in energy.

\section{Conclusion}

We have considered a series of electron bunch data obtained from Laser Plasma Accelerators at a various major 
experiments. Given the high peak current of such bunches we have investigated their evolution in simple transfer 
lines with the aim of assessing the effect of space charge forces. The results obtianed with CSR-Track show that 
the effect of space charge on the emittance and bunch length is generally negligible for the bunch data obtained 
experimentally. Of the previously published papers that were considered in this investigation, those with 
relatively large normalised emittance were found to be least affected. This was due to the fact that the initial 
divergence and emittance of the beams was too large for the Coulomb repulsion to have a serious impact.
However moving towards future generations of LPAs with smaller emittances and larger bunch charges as seen in 
the case of the OPALS, our investigations show that space charge will have a significant effect unless operational 
energies of the LPAs are increased.

\section{Acknowledgments}

The authors are grateful to the John Fell fund from the University of Oxford for the funding provided to support their work.

\end{document}